\documentclass[aps,pre,reprint,superscriptaddress,floatfix]{revtex4-1}

\usepackage{graphicx}
\usepackage{amsmath}
\usepackage{xcolor}
\usepackage[normalem]{ulem}
\usepackage{float}
\begin{document}

% Use the \preprint command to place your local institutional report
% number in the upper righthand corner of the title page in preprint mode.
% Multiple \preprint commands are allowed.
% Use the 'preprintnumbers' class option to override journal defaults
% to display numbers if necessary
%\preprint{}

%Title of paper
\title{Synchronization in bus systems with partially overlapping routes}

% repeat the \author .. \affiliation  etc. as needed
% \email, \thanks, \homepage, \altaffiliation all apply to the current
% author. Explanatory text should go in the []'s, actual e-mail
% address or url should go in the {}'s for \email and \homepage.
% Please use the appropriate macro foreach each type of information

% \affiliation command applies to all authors since the last
% \affiliation command. The \affiliation command should follow the
% other information
% \affiliation can be followed by \email, \homepage, \thanks as well.
%==============================
\author{Sakurako Tanida}
\email[]{u-tanida@g.ecc.u-tokyo.ac.jp}
%\homepage[]{Your web page}
%\thanks{}
\affiliation{Department of Aeronautics and Astronautics, School of Engineering, The University of Tokyo, 7-3-1 Hongo, Bunkyo-ku, Tokyo, 113-8656, Japan}
\affiliation{Institute for Multiscale Simulation, Friedrich-Alexander-Universit\"at Erlangen-N\"urnberg, Cauerstra{\ss}e 3, D-91058 Erlangen, Germany}
%==============================
\author{Thorsten P\"oschel}
%\email[]{Your e-mail address}
%\homepage[]{Your web page}
%\thanks{}
%\altaffiliation{}
\affiliation{Institute for Multiscale Simulation, Friedrich-Alexander-Universit\"at Erlangen-N\"urnberg, Cauerstra{\ss}e 3, D-91058 Erlangen, Germany}
%==============================

%Collaboration name if desired (requires use of superscriptaddress
%option in \documentclass). \noaffiliation is required (may also be
%used with the \author command).
%\collaboration can be followed by \email, \homepage, \thanks as well.
%\collaboration{}
%\noaffiliation

\date{\today}

\begin{abstract}
In an increasingly interconnected world, understanding congestion-related phenomena in transportation and their underlying mechanisms is crucial for improving efficiency.
As the transportation system becomes denser, different modes of transportation have more opportunities to interact with each other, giving rise to emergent dynamics that simple models cannot explain.
In this study, we investigate the synchronized motion of indirectly coupled transportation modes.
We develop a numerical simulation model on a one-dimensional periodic lattice, where each point represents a bus station. In this system, two types of buses operate: multiple local buses with non-overlapping routes, each serving a specific zone, and a single global bus that partially overlaps with the routes of the local buses.
We perform numerical simulations to examine how close the arrival times of these buses are to each other---that is, how synchronized their motions are.
When the number of zones is two, three, or five, robust synchronization occurs not only between the global bus and the local buses, but also among the local buses themselves. In contrast, no synchronization is found for other numbers of zones. We developed a mathematical model using self-consistent equations and found that two distinct arrival patterns at the terminals must be considered. A stability analysis reveals which pattern is ultimately realized in the simulations.
Our results show that transportation modes can exhibit coherent motion even when sharing only partial or no direct route overlaps. This outcome highlights that emergent behavior depends not only on local interactions but is also strongly shaped by the system’s overall structural configuration.
\end{abstract}

% insert suggested keywords - APS authors don't need to do this
%\keywords{}

%\maketitle must follow title, authors, abstract, and keywords
\maketitle

% body of paper here - Use proper section commands
% References should be done using the \cite, \ref, and \label commands
%\section{}
% Put \label in argument of \section for cross-referencing
%\section{\label{}}
%\subsection{}
%\subsubsection{}
\section{Introduction}
The increasing demand for efficient transportation systems, driven by global population growth and globalization, has become increasingly evident. Simultaneously, technological advancements enable the development of new and sophisticated transportation systems. As a result, all modes of transportation are interconnected, with the dynamics of one system influencing others, even when they are only weakly or indirectly connected. This interconnectedness adds a layer of complexity that requires understanding the underlying mechanisms governing these systems before effective optimization strategies can be developed. 

Because transportation systems inherently involve energy dissipation through vehicle movement and interactions, they can be naturally regarded as non-equilibrium dissipative systems driven by energy inputs \cite{Nagatani2002, Helbing2001, Chowdhury2000}. This perspective provides a fundamental basis for understanding their macroscopic behaviors. For example, studies on how short headways between vehicles trigger deceleration leading to traffic jams \cite{Musha1976, Musha1978, Kerner1993, Komatsu1995, Kurtze1995} have elucidated various efficiency-related phenomena \cite{Hall1986, Neubert1999}. 

Another important example involves transportation modes that operate on a periodic trajectory, commonly found in public transportation, such as buses, trains, and elevators. These modes can therefore be modeled as oscillators \cite{Poschel1994}. These transportation modes exhibit bunching, superficially resembling traffic jams. However, this phenomenon is primarily driven by a positive feedback loop via waiting passengers. When the distance between a leading bus and a following bus becomes shorter, both the probability and the number of passengers waiting at stations ahead of the following bus decrease, effectively increasing its average speed over time \cite{Oloan1998}. 

This oscillator-based viewpoint naturally raises questions about synchronization, namely whether it can extend to broader, network-wide effects, or whether route fragmentation prevents global coordination. Such route configurations, in which multiple transportation modes share only certain segments while remaining connected through the overall network structure, are commonly observed in real transportation systems. From a practical perspective, therefore, it is important to understand how localized interactions might (or might not) scale to broader network dynamics. At first glance, synchronization appears likely if oscillators interact for a sufficiently long time within a shared segment. However, once an oscillator interacts with another oscillator in an adjacent shared segment, it remains unclear whether synchronization established in the original segment will remain stable in the adjacent segment.

In this study, to clarify how partial interactions affect global order, we investigate a transportation system in which local buses serve only stations within specific zones. In contrast, a global bus traverses multiple zones, providing service across their boundaries. By simulating this setup with a discrete model, we examine the order parameters for local buses that do not directly interact, as well as for local–global buses, under different parameter sets---specifically, the number of zones ($N$) and the passenger inflow rate ($\mu$). Our results show that the buses can synchronize their motions depending on $N$: for $N=2,3$, and 5, a high degree of order is observed, whereas other values of $N$ do not yield such an order. When examining the ratio of the local buses' round-trip time to the global bus's cycle time, we find that a high degree of order emerges only if this ratio is an integer. To explain this dependence on the parameter set, we developed a mathematical model. By considering the arrival pattern in which either the local or the global bus arrives at the terminal first, we can calculate the typical time each bus spends at every station, from which the period ratio can be estimated.

In what follows, we formulate the problem by specifying the system's setup and key parameters (Section~\ref{sec:pf}). We then present our results (Section~\ref{sec:results}), beginning with the case $N=2$, where we conduct both numerical simulations and modeling analysis. We then extend the same approach to larger $N$, revealing how the synchronization behavior depends on the system structure. Our discussion (Section~\ref{sec:dis}) explains why synchronization emerges only for specific $N$ values and considers additional factors that may influence these outcomes. Finally, the conclusion (Section~\ref{sec:con}) summarizes our main findings and discusses potential directions for further research.

\section{Problem Formulation}\label{sec:pf}
\begin{figure}[h]
\centering
\includegraphics[width=0.8\linewidth]{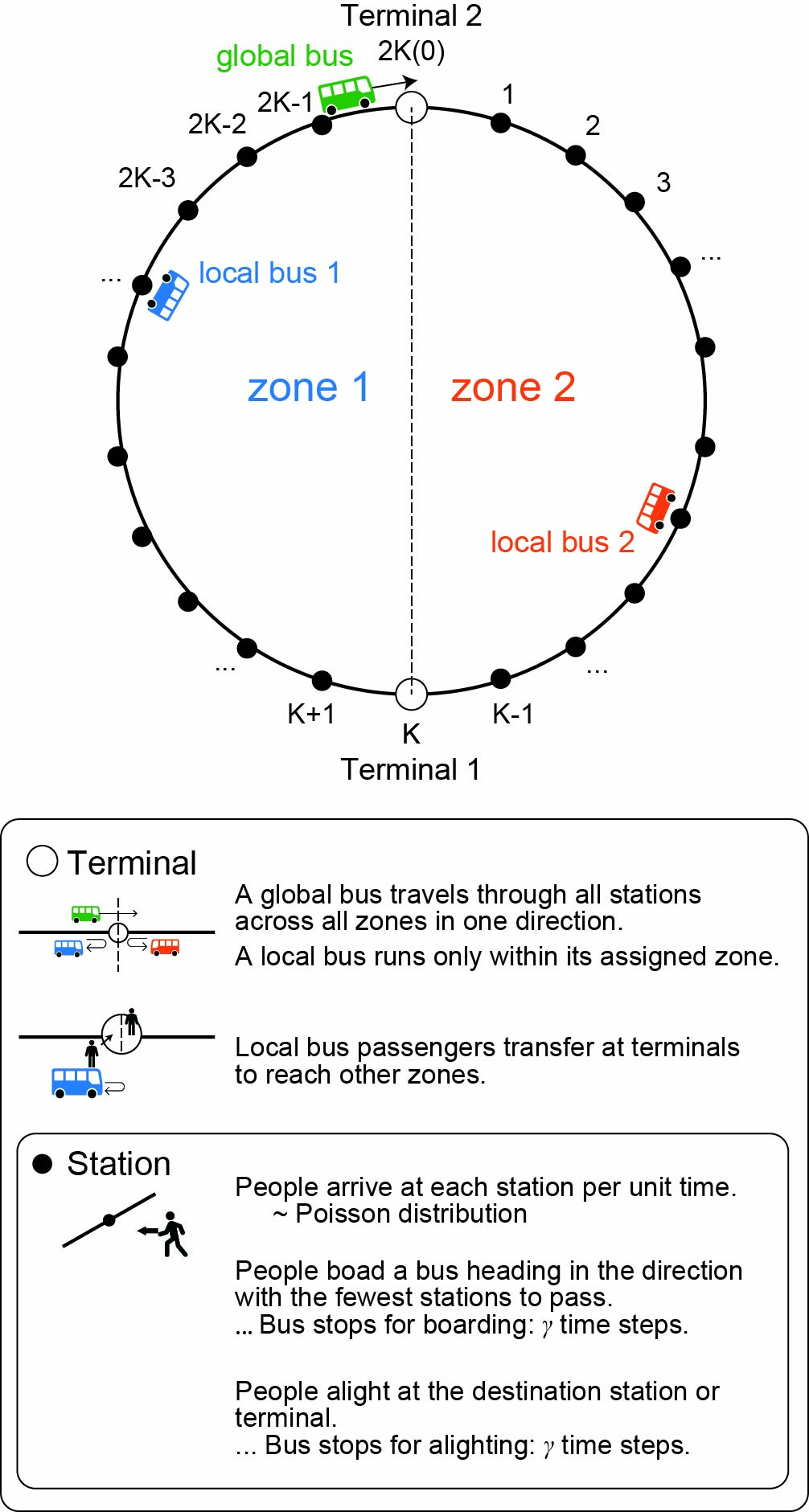}
\caption{
Schematic and summary table of the formulated problem. The upside shows a one-dimensional system with periodic boundary conditions, illustrating an example with two zones. Each zone is serviced by a local bus (blue for zone 1 and orange for zone 2), while the global bus (green) traverses both zones continuously without changing direction. The system operates on a circular route, with stations represented by black circles and labeled by number.
The bottom side summarizes bus and passenger behaviors at both terminals and regular stations.}
\label{fig:setup}
\end{figure}

Our model system is related to the Bus Route Model (BRM) \cite{Oloan1998}; each lattice point in the periodic one-dimensional space corresponds to a bus station (Fig.~\ref{fig:setup}). The unit of time (also referred to as \textit{time step}) is defined by the time required by the bus to progress to the next station. We consider bus stopping for both boarding and alighting. If a bus arrives at a station with waiting passengers, it stops for $\gamma$ time steps for boarding. When passengers on board alight, the bus stops for another $\gamma$ time steps. In this study, we fixed $\gamma=10$.

%==FIG============================
%==FIG============================

Buses are allowed to overtake other buses. Specifically, overtaking may occur if a second bus arrives at a station where another bus is currently stopped, and the second bus has no passengers needing to alight. In that case, the second one simply passes through without stopping.

The main difference between this model and the BRM~\cite{Oloan1998} is that the system is divided into $N$ zones, each containing $K$ stations, resulting in a total of $NK$ stations (Fig.~\ref{fig:setup}). Stations with labels $nK$ for $n \in \{1,2, \dots, N\}$ are shared between zones $n-1$ and $n$, and are called terminals. As an exception, a station with labels $NK$ is shared between zones $N$ and 1 due to periodic boundary conditions. Local buses are restricted to serving stations within their designated zone; a local bus in zone $n$ serves stations labeled from $(n{-}1)K$ to $nK$. At terminals, they change direction to avoid entering other zones. In contrast, the global bus serves all $NK$ stations and maintains a constant direction of motion without changing direction. In this study, we fixed $K=20$.

The number of passengers that appear at each station to enter a bus follows a Poisson process with parameter $\mu/K$. Specifically, the probability that $n$ passengers arrive at a given station within a time step is given by:
\begin{equation}
P(n) = \frac{\lambda^n e^{-\lambda}}{n!},
\end{equation}
where $\lambda = \mu/K$ represents the average arrival rate per station.
Each passenger has an individual destination and waits for a bus heading in the direction that minimizes the number of stations to pass. If the destination is located in a different zone, the passenger must change buses at a terminal and wait for the local bus serving the next zone or for the global bus.

\section{Results}\label{sec:results}
\subsection{Two-zone case ($N=2$)}\label{sec:n2}
\begin{figure*}[h]
    \centering
    \includegraphics[width=0.99\linewidth]{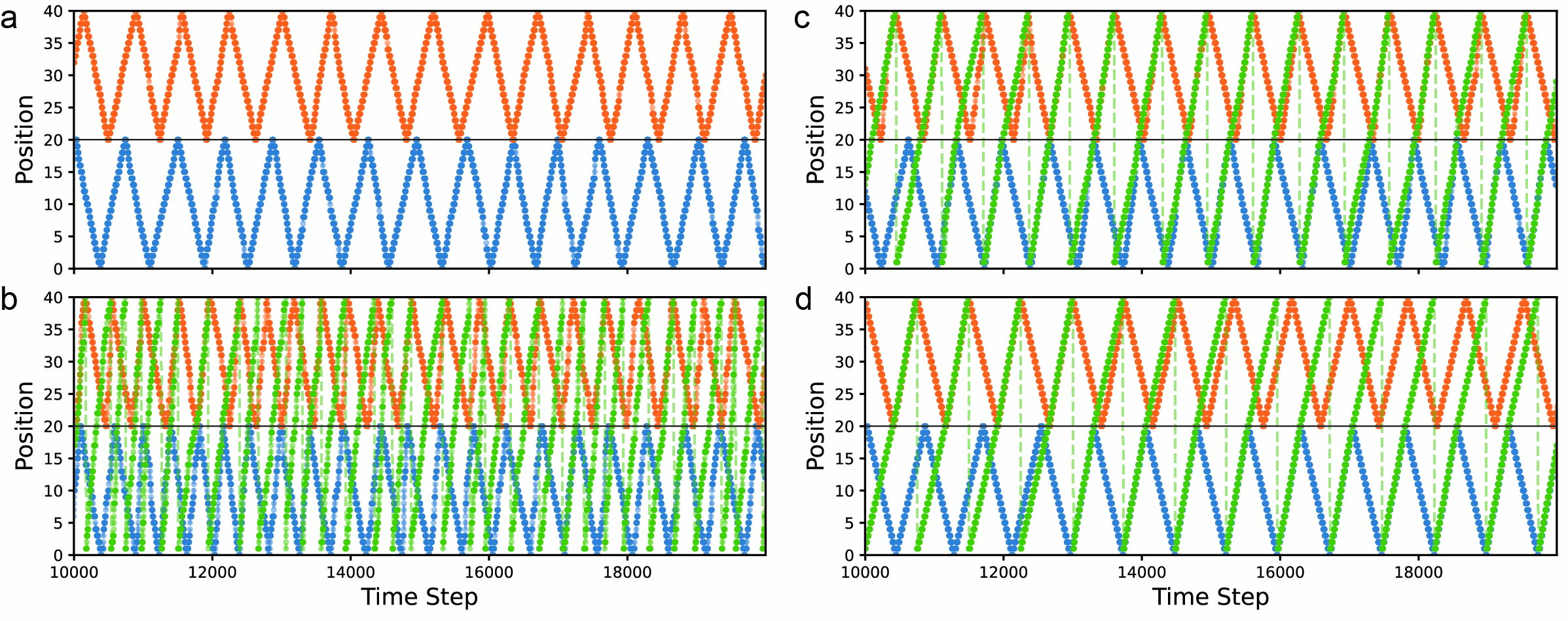}
    \caption{
    Positions of the buses as functions of time, for $N=2$. (a) Without the global bus and with $\mu=0.1$, the average round-trip time of the local buses is 760 time steps. (b) With the global bus and with $\mu=0.1$, their average round-trip time is 546 time steps. (c) With the global bus and $\mu=0.2$, the round-trip time is 752 time steps, close to the value in (a). (d) With the global bus and $\mu=0.7$, the round-trip time increases to 812 time steps, approaching the scenario where the local buses stop at every station. In each panel, different colors represent different buses: the orange and blue lines correspond to local buses, while the green line represents the global bus.
    }
    \label{fig:exampleN2}
\end{figure*}

We first examine the behavior for two zones ($N=2$) without a global bus. Figure~\ref{fig:exampleN2}a shows the position of the local buses for the Poisson parameter, $\mu=0.1$. The local buses exhibit a characteristic round-trip time, $T_L$, defined as the duration from when a local bus departs a terminal until it departs from the same terminal again. Despite this, the arrival times at the terminals are stochastic. That is, no synchronization is found.

%==FIG============================
%==FIG============================

\begin{figure}[bt]
\centering
\includegraphics[width=0.99\linewidth]{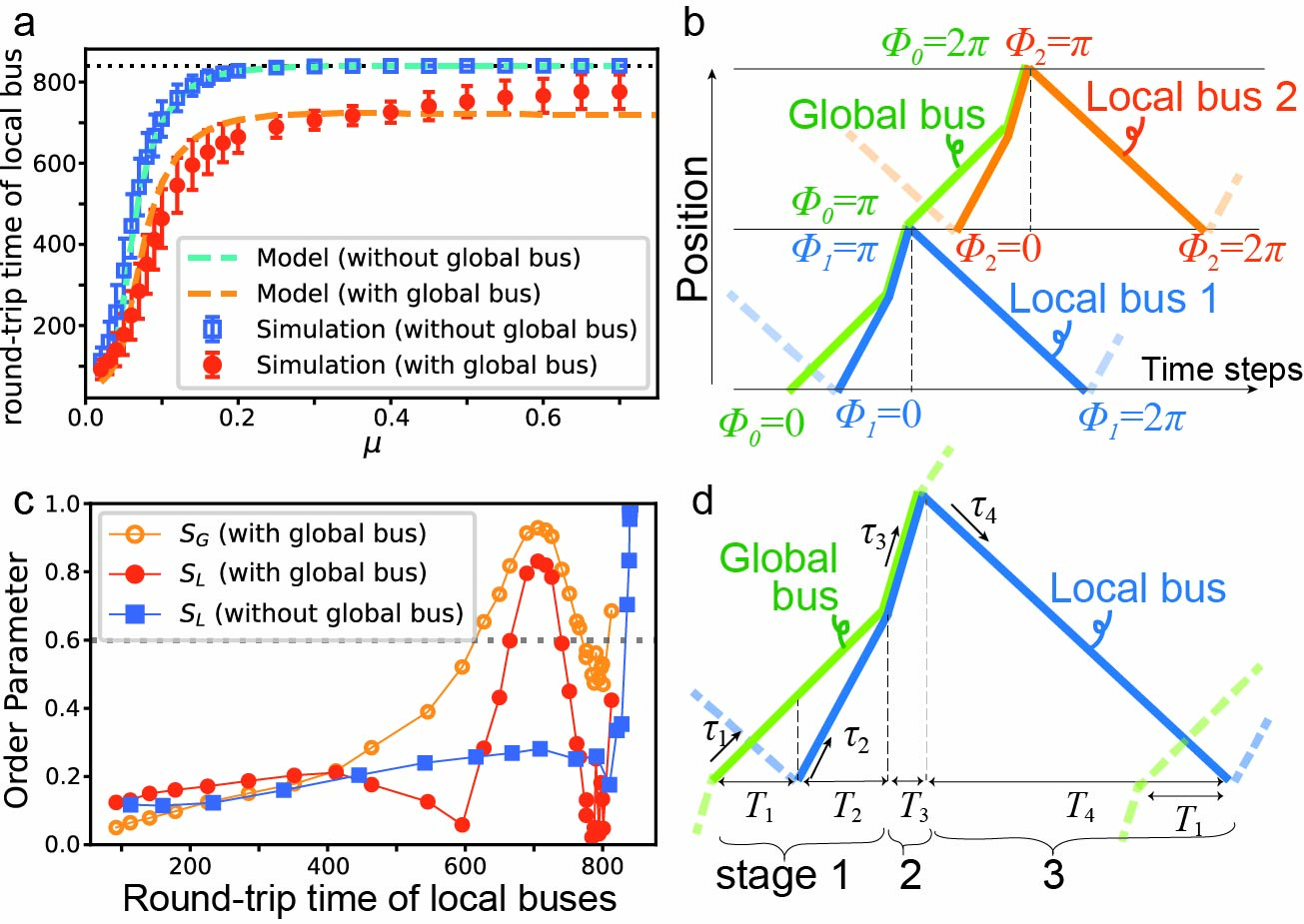}
\caption{
(a) Average round-trip time as a function of passenger inflow rate $ \mu $ for systems with and without a global bus. Error bars indicate one standard deviation. Cyan and orange dashed lines show the theoretical values with and without a global bus, respectively, obtained by solving the self-consistent equations. The black dotted line represents the maximum round-trip time of the local bus.
(b) Illustration of how the phases of local and global buses are defined.
(c) Averaged synchronization parameters, $ S_G $ and $ S_L $, as functions of the round-trip time of local bus. The dashed lines serve as visual guides to help compare whether the peak values exceed 0.6 across different values of $N$.%The reason why $ S_L $ without a global bus becomes high when the round-trip time of the local buses exceeds 800 is due to an artifact: buses stop at nearly all stations, so the distances between them remain almost unchanged during the simulation, leading to a high order parameter.
(d) Sketch of the mathematical model for the GBL scenario, in which the periodic synchronization pattern is divided into three stages.
}
\label{fig:N2order}
\end{figure}

To understand how the round-trip time behaves more generally, we now examine the dependence on $\mu$. A large standard deviation of the round-trip time can be found for small $\mu$ when the number of passengers is small. The average round-trip time increases with $\mu$, as shown with blue squares in Fig.~\ref{fig:N2order}a. For $\mu \gtrsim 0.2$, the round-trip time converges toward a maximum value, since the number of passengers is so large that each bus stops at every station. In this regime, the round-trip time is nearly constant, and the standard deviation is smaller than that for lower $\mu$.

For the system of two local buses, the average round-trip time can be obtained from a self-consistent equation similar to those addressed in \cite{Tanida2021}. The probability $p(t)$ that passengers board at each floor after time $t$, measured from the last arrival in the direction, is given by:
\begin{eqnarray}
p(t) &=& 1 - \exp\left(-\frac{\mu t}{2K}\right). \label{eq:sc}
\end{eqnarray}
Note that the Poisson parameter $\mu$ enters the exponent with the factor 1/2 to account for passengers choosing randomly between two directions. After a local bus completes its round trip of duration $T_L$, the probability that passengers will need to board upon the bus’s return is given by $p(T_L)$; the same form is applied for alighting. Thus, the round-trip time can be calculated from the expected number of floors where the bus stops in each direction, which is $2K \cdot p(T_L)$. This leads to the round-trip time
\begin{eqnarray}
T_L = 2K + 4\gamma K p(T_L) \,.
\label{eq:basic}
\end{eqnarray}
The average period $T_L$ is the solution of this implicit equation shown in Fig.~\ref{fig:N2order}d (cyan line). This analytical solution is in very good agreement with the result obtained from the simulation, shown as blue squares in Fig.~\ref{fig:N2order}d. Defining the average time spent at each station by $T^\prime = {T_L}/{K}$, Eq.~\eqref{eq:basic} reads 
\begin{eqnarray} 
 T^\prime = 2 + 4\gamma \left[1 - \exp\left(-\frac{\mu T^\prime}{2}\right)\right], \label{eq:Kindependent}
\end{eqnarray} 
which shows that the average time spent at each station is independent of the total number of stations, $K$.

We now consider the same two-zone system as before but with a global bus. Figure~\ref{fig:exampleN2}c shows the positions of the buses as functions of time for $\mu = 0.2$. We see that the average round-trip time for local buses is similar to that of the previous case, without the global bus as shown in Fig.~\ref{fig:exampleN2}a. However, the arrival times of the buses at the terminals are synchronized because of the presence of the global bus---specifically when they approach the terminal from the same direction as the global bus. In the opposite direction, where the global bus is absent, simultaneous arrival does not occur. For low inflow, $\mu=0.1$ (Fig.~\ref{fig:exampleN2}b), the typical round-trip time is noticeably shorter, and synchronization is suppressed. This is plausible because, for $\mu=0$, the buses effectively run independently, as there are no passengers that would cause them to stop at any station. Synchronization is also suppressed for high inflow, $\mu=0.7$, (Fig.~\ref{fig:exampleN2}d). The similar principle applies when $\mu$ becomes very large: each bus is forced to stop at nearly every station, effectively removing any interactive influence between the buses and thus suppressing synchronization.

The average round-trip time is significantly shorter in the system with the global bus than without, for all values of $\mu$, as shown in Fig.~\ref{fig:N2order}a. This is because the global bus serves some stations, so local buses can omit certain stations for boarding and alighting. For sufficiently large $\mu$, the arrival times of local buses in the system without the global bus become regular because they stop at nearly every station; in contrast, the arrival times in the system with the global bus are regular due to the interaction with the global bus. To properly distinguish how the global bus affects arrival times from the impact of passenger inflow rates, it is more appropriate to compare cases with the same round-trip time, such as Figs.~\ref{fig:exampleN2}a and \ref{fig:exampleN2}c, rather than cases with the same $\mu$, as in Figs.~\ref{fig:exampleN2}a and \ref{fig:exampleN2}b.

To quantify the degree of synchronization, we define the phase of the local bus in zone $n$ located at station $k$ at time $t$ by 
\begin{equation}
\phi_{n,t} =
\begin{cases}
    \frac{k-(n-1)K}{K}\pi & \text{for~~} v_n>0 \\
    \frac{(n+1)K-k}{K}\pi & \text{for~~} v_n<0
\end{cases}
\end{equation}
where $v_n$ denotes the direction of motion; $v_n > 0$ indicates the motion in increasing station number, and $v_n < 0$ in decreasing.
Similarly, the phase of the global bus at station $k$ at time $t$ is defined by
\begin{equation}
\phi_{\,0,t} = \frac{2 \pi k }{KN}  \ . \label{eq:phi0}
\end{equation}

With these definitions, we introduce two order parameters,
\begin{eqnarray}
    S_G &\equiv&  \frac{1}{N}\sum_{n=1}^{N} \left|\frac{1}{T}\sum_{t=0}^T\exp \left( i \left[\phi_{n,t} - \phi_{0,t} \right] \right)\right| \label{eq:sg}\\
    S_L &\equiv& \frac{1}{N-1}\sum_{n=1}^{N-1} \left|\frac{1}{T}\sum_{t=0}^T \exp \left( i \left[\phi_{n,t} - \phi_{n+1,t} \right] \right) \right|
\end{eqnarray}
to quantify the alignment of the local buses with the global bus ($S_G$) and the synchronization between consecutive local buses ($S_L$).

Figure \ref{fig:N2order}c shows the average across ten samples of $S_L$ and $S_G$ for the system containing a global bus and for the system without a global bus. The $x$-axis represents the average round-trip time of the local buses. The value of $S_L$ without the global bus remains around 0.2; note that it shows a sudden increase when the round-trip time approaches approximately $2K + 4\gamma K = 840$. This is because the local buses need to stop at almost every station, preventing the initial positional relationships from relaxing into randomness. 

In comparison, the case with the global bus behaves differently. $S_L$ is significantly larger for the round-trip times between 620 and 780 time steps. Additionally, $S_G$ increases along with $S_L$, indicating that synchronization between the global and local buses is closely tied to synchronization and desynchronization between the local buses. Outside this period range, $S_L$ with the global bus is lower than without the global bus, even when $S_G$ remains high. This suggests that there is a mode where the local buses fail to synchronize with each other despite remaining synchronized with the global bus.

%==FIG============================
%==FIG============================

To understand the synchronization mechanism, we subdivide the motion into three stages (Fig.~\ref{fig:exampleN2}). 
At each stage, we treat the expected stopping time at each station as a continuous variable. The probability that a bus stops at a station depends on the time interval, defined as the elapsed time since the most recent arrival of any bus at that station.

The time the global bus spends at each station in stage 1, denoted as $\tau_1$ is given by:
\begin{equation} 
 \tau_1 = \tau(T_2 + T_3 + T_4-T_1), \label{eq:GL1}
\end{equation}
where $\tau(t)$ represents the expected time spent at a station, accounting for four possible scenarios: no boarding or alighting, boarding, alighting, or both. The function $\tau(t)$ is defined as:
\begin{eqnarray} 
 \tau(t) &=& \left[1 - p(t)\right]^2 + 2(\gamma + 1)p(t)\left[1 - p(t)\right] \nonumber \\
 &&+ (2\gamma + 1)p(t)^2 \ . \label{eq:tau}
\end{eqnarray}
We assume that the probability of alighting is the same as that of boarding. During the time $T_1$---from the global bus departure until the local bus departure---the global bus covers $x = T_1/\tau_1$ stations. With the interval $T_1$, the local bus spends $\tau_2 = \tau(T_1)$ at each station. Therefore, the time $T_2$ it takes for the local bus to catch up to the global bus is given by:
\begin{eqnarray} 
T_2 &=& x\left(\tau_1 - \tau_2\right). \label{eq:N2T2}
\end{eqnarray}

In stage 2, we assume that each bus skips boarding at half of the stations with passengers, as the other half are served by the other bus. Thus, the boarding probability at each station is halved, becoming $\frac{1}{2} p(t)$. However, the probability of stopping for alighting remains $p(t)$, unchanged. The expected time spent at a station, denoted as $\hat{\tau}(t)$, accounts for these adjusted probabilities. Specifically, $\hat{\tau}(t)$ is defined as:
\begin{eqnarray} 
 \hat{\tau}(t) &=& \left[1 - p(t)\right]\left[1 - \dfrac{1}{2}p(t)\right] \nonumber \\
 &&+ (\gamma + 1)\left\{\dfrac{1}{2}p(t)\left[1 - p(t)\right] + p(t)\left[1 - \dfrac{1}{2}p(t)\right]\right\} \nonumber \\&&+ \dfrac{2\gamma + 1}{2} p(t)^2, \label{eq:tauhat}
\end{eqnarray}
The time spent at a station by both buses is given by:
\begin{eqnarray} 
\tau_3 &=& \hat{\tau}(T_2 + T_3 + T_4) . \label{eq:N2tau3}
\end{eqnarray}
Here, the interval $T_2 + T_3 + T_4$ reflects the time since the last bus served the station. Finally, the duration $T_3$ of  stage 2 can be calculated as:
\begin{eqnarray} 
T_3 &=& \left(K + 1 - \dfrac{T_1 + T_2}{\tau_1}\right) \tau_3, \label{eq:N2T3}
\end{eqnarray}
where $K + 1$ is the total number of stations.

In stage 3, only the local bus serves the stations on its return trip. The time per station and the duration of this stage are:
\begin{eqnarray}\tau_4 &=& \tau(T_2 + T_3 + T_4), \\
 T_4 &=& (K + 1) \tau_4. \label{eq:N2cond}
\end{eqnarray}

Considering that the local bus makes a full round trip while the global bus makes a full cycle, we have:
\begin{eqnarray} 
N(T_1 +T_2 + T_3) &=&  T_2 + T_3 + T_4. \label{eq:GLreq}
\end{eqnarray}

These equations are self-consistent. Given $T_1$ and $T_2 + T_3 + T_4$, we can calculate $\tau_1$, $\tau_2$, $\tau_3$, and $\tau_4$. From these, we determine $T_2$, $T_3$, and $T_4$, which must sum to match the initially known value of $T_2 + T_3 + T_4$. The orange dashed line in Fig.~\ref{fig:N2order}a shows the round-trip time of the local buses, $T_2 + T_3 + T_4$, obtained from this procedure, showing good agreement with the simulation results.

According to the mathematical models, we can explain why the pattern with a global bus remains stable when small fluctuations are added to the interval times. In the system without a global bus, a slight change in the round-trip time to $T_L + \delta$ does not affect the next round trip significantly. In contrast, in the system with a global bus, substituting $T_1 + \delta$ into the self-consistent equations reveals that the subsequent interval becomes shorter. This means that the presence of the global bus stabilizes the interval and timing, leading to higher order in the system. The normalized round-trip time of the local bus, as obtained from Eqs.~\eqref{eq:GL1}–\eqref{eq:GLreq}, is independent of $K$, similar to the result shown in Eq.~\eqref{eq:Kindependent}.

\subsection{Multi-zone cases ($N>2$)}\label{sec:n3-6}
\begin{figure*}[ht]
\centering
\includegraphics[width=1\linewidth]{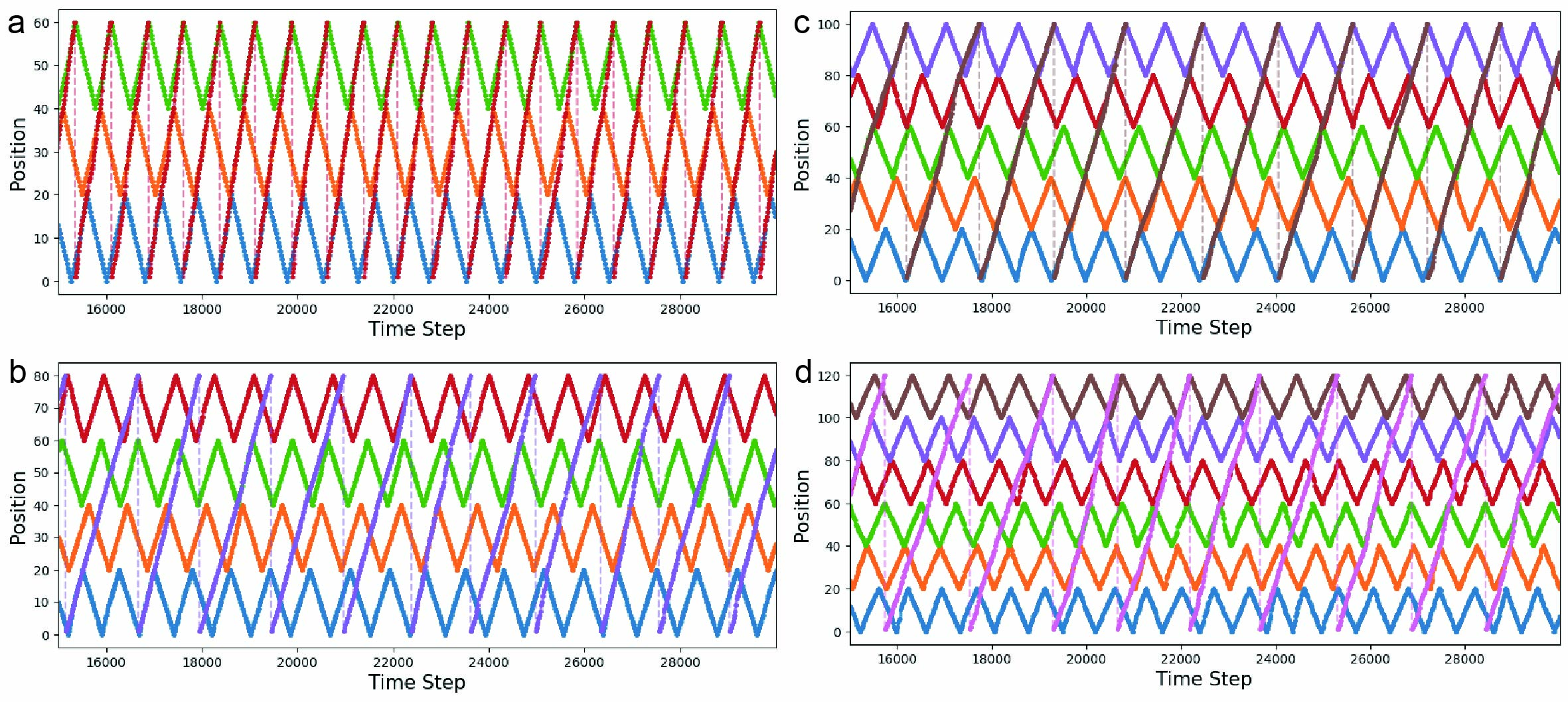}
\caption{
Time evolution of bus positions (a) for $N=3$ and $\mu=0.2$, (b) for $N=4$ and $\mu=0.25$, (c) for $N=5$ and $\mu=0.3$, and (d) for $N=6$ and $\mu=0.2$. In each panel, different trajectories represent different buses, with the global bus identifiable as the one traveling across all stations, while local buses operate within their respective zones. In (a) and (c), the global bus’s arrival at the terminals aligns with that of the local buses, whereas this simultaneous arrival pattern is not observed in (b) and (d). 
}
\label{fig:exampleN3-6}
\end{figure*}

Next, we investigate cases where the number of zones ($N$) is more than two. Figures \ref{fig:exampleN3-6}a–d show examples of the time evolution of bus positions for $N=3$, 4, 5, and 6. These examples were specifically chosen to represent cases where the order parameter reaches its highest value across various $\mu$. For $N=3$ and $N=5$, the global bus almost always arrives at each terminal simultaneously with the local buses, whereas for $N=4$ and $N=6$, this synchronization is not observed. For larger numbers of zones, the systems do not exhibit clear periodic synchronization patterns (no data shown), suggesting that synchronization becomes increasingly difficult as $N$ increases. Furthermore, as $N$ increases (from $N=2$), the number of local-bus round trips per global bus cycle increases.

%==FIG============================
%==FIG============================

\begin{figure}[ht]
\centering
\includegraphics[width=1\linewidth]{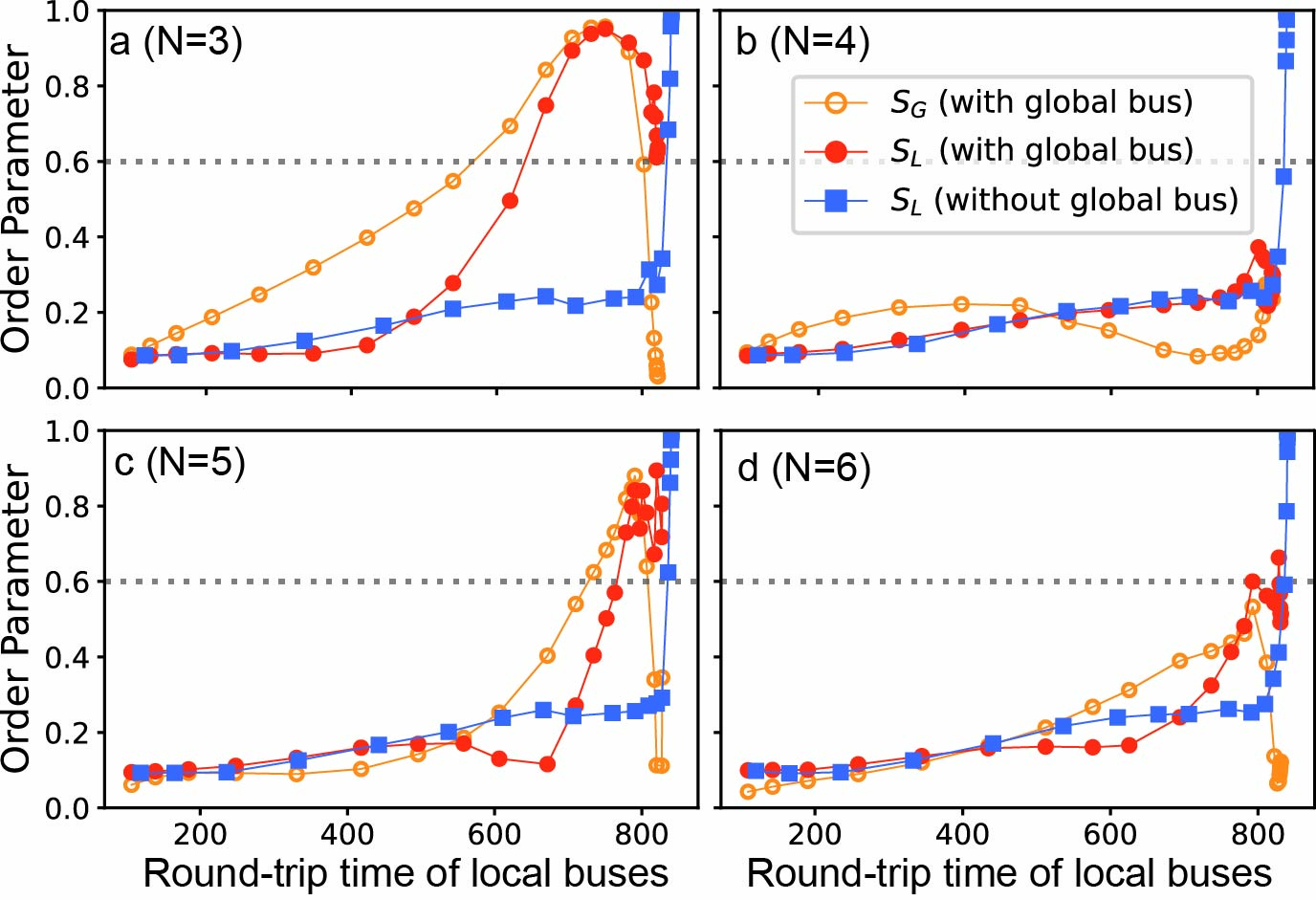}
\caption{The order parameters at (a) $N=3$, (b) $N=4$, (c) $N=5$, and (d) $N=6$. Blue squares represent $S_L$ without the global bus, red circles represent $S_L$ with the global bus, and orange circles represent $S_G$ with the global bus. The dashed lines serve as visual guides to help compare whether the peak values exceed 0.6 across different values of $N$.%The reason why $S_L$ without a global bus is high when the round-trip time of the local buses exceeds 800 is due to an artifact. When the round-trip time exceeds 800, the buses stop at almost all stations, and the distances between buses hardly change during the simulation time steps, resulting in a high order parameter.
}
\label{fig:orderN3-6}
\end{figure}

This tendency is consistent across various values of $\mu$. Figures \ref{fig:orderN3-6}a–d present the order parameter $S_L$ and $S_G$, averaged over at least ten independent simulations.
Note that the definition of $S_G$ follows Eq.~(\ref{eq:sg}), but here we adopt a modified form of the phase variable $\phi_{o,t}$, to reflect the case where multiple round trips of local buses occur in a single cycle of the global bus. In this section, we use the generalized form:
\begin{equation}
    \phi_{\,0,t} = \left( \frac{2 \pi k m}{KN}\right) \mod~~ 2\pi \ , \label{eq:phi0m}
\end{equation}
where $m$ is the period ratio, computed by
\begin{equation}
    m = \frac{T_G}{T_L}\,
\end{equation}
with $T_G$ defined as the average cycle time of the global bus (the duration from when it departs a terminal until it departs that same terminal again), and $T_L$ defined as the round-trip time of the local buses.
In the case of $N=2$, $m$ is effectively $1$, so Eq.~(\ref{eq:phi0m}) reduces to Eq.~(\ref{eq:phi0}).

In the absence of a global bus, $S_L$ remains low across different values of $N$, excluding the artifact observed at round-trip times exceeding 800, as indicated by the blue markers. However, when the global bus is introduced, $S_L$ shows a distinct increase in certain regions. For $N=3$, both $S_L$ and $S_G$ with the global bus are significantly higher than $S_L$ without the global bus for round-trip times over 500 time steps, but they drop outside this region, similar to the behavior observed for $N=2$. For $N=4, 5$, and 6, $S_L$ with the global bus increases in the long-period region, but the peak values differ: $N=5$ reaches a peak comparable to $N=3$, while $N=4$ and $N=6$ exhibit lower peaks. The trend of $S_G$ generally follows that of $S_L$ with the global bus, increasing and decreasing in a similar manner, except for $N=4$.

%==FIG============================
%==FIG============================

\begin{figure}[ht]
\centering
\includegraphics[width=1\linewidth]{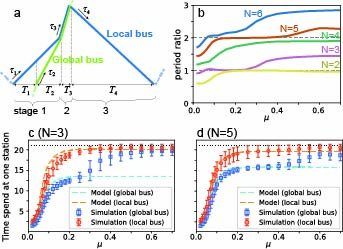}
\caption{
(a) Schematic of the mathematical model of the LBL scenario, where the repeated pattern is divided into three stages, defining each interval and the time spent at each station during each stage. 
(b) The period ratio for various values of $\mu$ and $N$. The period ratio is defined as the ratio of the global bus cycle time to the local bus round-trip time.
(c, d) The time spent at one station, averaged over the round trip or cycle, for various values of $\mu$ for $N=3$ and $N=5$, respectively. Blue squares represent the average of simulation values without the global bus, while red circles represent the average with the global bus. Cyan and orange dashed lines show the theoretical values obtained by solving the self-consistent equations, with and without the global bus, respectively. The black dotted line indicates the maximum time spent at one station.
}
\label{fig:intervalN35}
\end{figure}

The emergence of order appears to be closely related to the relative periods of the global and local buses, expressed as the ratio $T_L:T_G$. In this study, synchronization is observed when these periods form integer ratios (e.g., $1:1$, $1:2$). As shown in Fig.~\ref{fig:intervalN35}b, the period ratio of local bus round trips during one global bus cycle ($m$) tends to increase as $N$ increases. This trend suggests that, for large $N$, the influence of the global bus diminishes, and local buses behave nearly independently. For $N=2$, 3, and 5, the period ratio exhibits plateaus at integer values (e.g., 1 or 2) over a range of $\mu$. These plateaus suggest period entrainment, with the buses adjusting their periods through interaction. In contrast, for $N=4$ and $N=6$, such integer plateaus are not observed, implying that the range of 
$\mu$ considered does not permit the buses to spontaneously adjust their periods to integer ratios.

%==FIG============================
%==FIG============================

To further investigate which values of $N$ allow synchronization, we develop a mathematical model to determine the intervals between departures from the terminal. We consider two scenarios---global bus leading (GBL) and local bus leading (LBL)---both of which follow the same three-stage dynamics.

In the GBL scenario, as shown in the case of $N=2$ (Fig.~\ref{fig:N2order}d), the global bus departs first, catches up with the local bus (stage 1), they travel together to the opposite terminal (stage 2), and the local bus returns alone (stage 3).
The only difference lies in the relationship between the periods of the global and local buses, Eq.~\eqref{eq:GL1}. For the case of $N>2$, we can consider multiple round trips during a single cycle of the global bus, leading to:
\begin{eqnarray} 
    N(T_1 +T_2 + T_3) &=& m( T_2 + T_3 + T_4), 
\end{eqnarray}
where $m$ is the period ratio, computed as
\begin{equation}
m = \frac{\tau_G}{\tau_L}\, \label{eq:m}
\end{equation}
with the average time spent at each station over the round trip ($\tau_L$) or cycle ($\tau_G$) given by:
\begin{equation}
    \tau_G = \frac{\tau_1 (T_1 + T_2) + \tau_3 T_3}{T_1 + T_2 + T_3}\quad\text{and}\quad \tau_L = \frac{\tau_2 T_2 + \tau_3 T_3 + \tau_4 T_4}{T_2 + T_3 + T_4}\,.
\end{equation}

In the LBL scenario, as shown in Fig.~\ref{fig:intervalN35}a, the local bus departs first, continues alone until the global bus catches up (stage 1), then both travel together to the opposite terminal (stage 2), and finally, the local bus returns alone in the opposite direction (stage 3).

The equations for the LBL scenario are almost the same as those for the GBL scenario, but the time spent at each station differs. $\tau_1$ and $\tau_2$ are switched for the global and local buses. Using the interval differences, $\tau_1$, $\tau_3$, and $\tau_4$ are given by:
\begin{eqnarray}
    \tau_1 &=& \tau(T_2 + T_3 + T_4 ), \\
    \tau_3 &=& \hat{\tau}(T_1+T_2 + T_3 + T_4), \\
    \tau_4 &=& \tau(T_1+T_2 + T_3 + T_4),
\end{eqnarray}
where $\tau(t)$ and $\hat{\tau}(t)$ are defined as in the GBL scenario in Eqs.\eqref{eq:tau} and \eqref{eq:tauhat}.
Assuming that both buses arrive simultaneously at the terminal, the following condition must hold:
\begin{eqnarray} 
    N(T_2 + T_3) &=& m(T_1+T_2 + T_3 + T_4), 
\end{eqnarray}
where $m$ represents the period ratio. In this scenario, $\tau_L$ and ($\tau_G$) are given by:
\begin{equation}
    \tau_G = \frac{\tau_2 T_2 + \tau_3 T_3}{T_2 + T_3}\quad\text{and}\quad \tau_L = \frac{\tau_1(T_1+T_2) + \tau_3 T_3 + \tau_4 T_4}{T_1+T_2 + T_3 + T_4}\,.
\end{equation}

The scenario that appears as a stable pattern in the system can be assessed through a stability analysis. In this analysis, a small fluctuation is introduced in the travel time, such that $T_1$ becomes $T_1 + \delta$. We then examine whether the next round-trip time returns to a value smaller than or equal to $T_1 + \delta$. If the fluctuation decays over subsequent trips, the system is stable; otherwise, the system is unstable, and the scenario does not persist. For all values of $\mu$ at $N=2$, the GBL scenario is stable, while the LBL scenario is unstable. In contrast, for $N=3$ and $N=5$, the LBL scenario is stable, and the GBL scenario is unstable.

Figures \ref{fig:intervalN35}c and \ref{fig:intervalN35}d show the average time spent at each station for both the global and local buses in each stable scenario. The cyan and orange dashed lines represent the theoretical values for the global and local buses in a stable scenario, $\tau_G$ and $\tau_L$, respectively. The theoretical values align well with the simulation results in the low $\mu$ region, particularly when the period ratio remains an integer and reaches its saturation values. However, for high $\mu$, the interval of the global bus in the simulation shifts to different values as the period ratio $m$ increases. This leads to scenarios that deviate from the initial assumptions.

\section{Discussion}\label{sec:dis}
In this study, we explored synchronization in which oscillators interact through partially shared phases, by investigating transportation systems featuring multiple local buses restricted to specific zones and a global bus that travels partially overlapping these zones. Although local buses remain uncorrelated on their own, introducing a global bus can induce synchronization, which is reminiscent of remote synchronization \cite{Bergner2012}, where elements synchronize through intermediary nodes without direct connections---but only under the specific number of zones; the synchronization arises only for the number of zones $N=2$, 3, and 5, while no synchronization appears for other values. 

This dependency on the number of zones can be explained by several points. First, the number of zones must remain small enough. As the number of zones increases, the global bus takes longer to pass through every zone, and local buses complete more round trips before the next interaction. This results in fewer opportunities for adjustment, which can lead to a breakdown in synchronization. 

Second, the ratio of the average round-trip times of the global and local buses governs synchronization; it occurs only when their period ratio is an integer. This integer relationship is observed over a wide range of passenger inflow rates, $\mu$, indicating that period entrainment arises as a result of interactions between the global and local buses. Notably, it emerges most clearly in the intermediate regime—unlike the low- and high-$\mu$ regimes where noise is minimal—indicating that the interaction between buses is strong enough to overcome the desynchronizing effects of fluctuations.
 
The mathematical models provided key insights into this phenomenon. First, consistent with previous studies \cite{Oloan1998,Poschel1994}, the interaction is driven by whether passengers are waiting at each station. Second, synchronization is stabilized by two distinct arrival patterns at the terminals: either the global bus or the local bus arrives first at each terminal. Finally, the models reveal that the system's qualitative behavior does not depend on the total number of stations, $K$, indicating that the mechanism holds regardless of the network’s overall size. However, this modeling approach assumes that the system reaches a stable arrival pattern and thus does not fully capture cases where the order remains low or no distinct pattern emerges, underscoring the need for further research to address those conditions.

Another aspect worth noting is that the passenger transfers at terminals, as modeled in this study, do not significantly affect the behavior of local buses in neighboring zones. Although buses from different zones converge at the same terminal, their routes are distinct, and they serve passengers traveling in separate directions. This separation ensures that interactions between zones remain minimal. Future work could explore how other factors, such as increased inter-zone interactions or alternative transfer mechanisms at terminals, might impact synchronization dynamics.

\section{Conclusion}\label{sec:con}
This study shows that transportation systems with fixed routes and designated stations, such as buses or elevators, can synchronize their motions even when they share only some or no stations. Furthermore, synchronization is influenced not only by the inflow rate of passengers at each station but also by the structure of the routes themselves. Specifically, systems with two, three, and five zone divisions showed clear synchronization, whereas those with any other number of zones did not exhibit the same degree of order. These findings provide a foundation for studying order formation in more complex transportation networks.

On the other hand, this study focused on a limited set of transportation system structures with rotational symmetry. Further research is needed to explore other relationships between buses and the underlying mechanisms that govern synchronization. Understanding how stable this synchronization is under more varied conditions, such as non-uniform inflow rates, will be critical for applying these findings to designing more efficient and resilient transportation infrastructures.

\begin{acknowledgments}
This work was supported by the JSPS Core-to-Core Program ``Advanced core-to-core network for the physics of self-organizing active matter (JPJSCCA20230002)''. We acknowledge support by the Interdisciplinary Center for Nanostructured Films (IZNF), the Competence Unit for Scientific Computing (CSC), and the Interdisciplinary Center for Functional Particle Systems (FPS) at Friedrich-Alexander University Erlangen-Nürnberg. The authors used ChatGPT, a language model developed by OpenAI, for minor language editing. The authors are solely responsible for the content of this work.
\end{acknowledgments}

\section*{Author Contributions}
TP proposed the problem and developed ideas for its analytical description, ST carried out the numerical simulations, did the analytical calculations and wrote the first draft of the paper. Both authors discussed the numerical data and their interpretation. Both authors revised the manuscript.
% Create the reference section using BibTeX:
%\bibliography{sample}

\end{document}